\documentclass[12pt]{article}
\usepackage{amssymb,amsmath,mathrsfs}
\usepackage{amsthm}
\usepackage{graphicx,epsfig}
\usepackage{caption}
\usepackage{subcaption}
\usepackage{enumerate}
\usepackage{epstopdf}
\usepackage{xcolor}
\usepackage{float}
\usepackage{rotating}
\usepackage{tikz}
\usepackage{dsfont}
\usepackage{comment} 
\usepackage{tabularx}
\usepackage{hyperref}
\usepackage{upgreek}
\usepackage[margin=3cm]{geometry}
\usepackage[english]{babel}
\usepackage[square,numbers]{natbib} 
\usepackage[linesnumbered,lined,boxed,commentsnumbered]{algorithm2e}
\RestyleAlgo{ruled}
\SetKwComment{Comment}{/* }{ */}
\hypersetup{
    colorlinks=true,
    linkcolor=cyan,
    filecolor=cyan,      
    urlcolor=cyan,
    citecolor = cyan
    }
\graphicspath{{Rips sequences/}}
 \numberwithin{equation}{section}

\DeclareUnicodeCharacter{0342}{\~{n}}
\DeclareUnicodeCharacter{0342}{\'{A}}

\begin{document}
\newcounter{eqnarray}
\newtheorem{t1}{Theorem}[section]
\newtheorem{d1}{Definition}[section]
\newtheorem{c1}{Corollary}[section]
\newtheorem{l1}{Lemma}[section] \newtheorem{r1}{Remark}[section]
\newtheorem{e1}{Example}[section]
\newtheorem{p1}{Proposition}[section]
\newtheorem{a1}{Result}[section]
\newtheorem{A1}{Assumption}

\title{Identifying Topological Differences in Two Populations of Random Geometric Objects} 
\author{Satish Kumar$^1$ and Subhra Sankar Dhar$^2$\\
Department of Mathematics and Statistics, Indian Institute of Technology Kanpur\\ 
Kanpur 208016, India\\
Emails: satsh@iitk.ac.in$^1$,
subhra@iitk.ac.in$^2$}
\maketitle
\begin{center}
    \textbf{Abstract} 
\end{center} 

We propose a statistical framework to identify topological differences in two populations of random geometric objects. The proposed framework involves first associating a topological signature with random geometric objects and then performing a two-sample test using the observed topological signatures. We associate persistence barcodes, a topological signature from topological data analysis, with each observed random geometric object. This, in turn, yields a two-sample problem on the space of persistence barcodes. As the space of persistence barcodes is not suitable for standard statistical analysis, we translate the two-sample problem on a suitable subset of a Euclidean space. In the course of this study, we embed the topological signatures in an ordered convex cone in a Euclidean space using functions from tropical geometry. We show that the embedding is a sufficient statistic for the persistence barcodes. This fact leads to the proposal of a two-sample test based on this sufficient statistic, and its equivalence to the two-sample problem on the barcode space is established. Finally, the consistency of the proposed test is studied.


\section{Introduction} \label{Introduction}
Topological data analysis (TDA) (see, e.g., \cite{Carlsson_Vejdemo-Johansson_2021}, \cite{Carlsson2020}, \cite{Chazal2021}, \cite{doi:10.1146/annurev-statistics-031017-100045} and references therein) is an emerging field that utilizes algebraic topological techniques to analyze complex and high-dimensional data. The foundation of TDA is laid on the so-called ``\emph{manifold hypothesis}" (\cite{Manifoldhypothesis}), which conjectures that high-dimensional data are sampled from a smooth manifold. The theme of TDA is that ``\emph{data has shape}",  and the shape (or the manifold) underlying the data may reveal stimulating insights about the process that generates the data, particularly when the data are high-dimensional and admit a complex structure.  

One of the key tools in TDA is persistent homology (see, e.g., \cite{Edelsbrunner(2008)}, \cite{Edelsbrunner2002}, \cite{ZomorodianCarlsson}), a multiscale extension of homology that is a classical topological invariant from algebraic topology (see, e.g., \cite{Munkres(1984)}, \cite{Hatcher(2002)}). Intuitively, homology characterizes a topological space using connected components and holes in higher dimensions by associating a sequence of abelian groups, called homology groups. Persistent homology is an adaptation of homology to sampled data points from geometric objects, where data are represented as finite metric spaces called \emph{point clouds}. Persistent homology summarizes topological features of data sets by associating a multi-set of intervals in real lines, called \emph{barcodes}. Barcodes provide a geometric and topological summary of the data-generating mechanism (see, e.g., \cite{PersistenceBarcodes}, \cite{barcode}, \cite{Ghrist(2007)}).

In the standard TDA framework, a point cloud is given from an unknown geometric object, and the goal is to infer the geometric and topological features of the underlying latent geometric object (see, e.g., \cite{Robinson2017}, \cite{Blumberg2014}, \cite{Fasy2014}, \cite{Tdaconsistency}, \cite{PDconvergence2014}, \cite{RobustCh(2017)}, \cite{chazal:GeometricInference-2009}). However, occasionally, we apply TDA tools to a different framework, where we consider a random sample of geometric objects instead of a point cloud sampled from an unknown geometric object. The aim here is to provide statistical inference on the probability distribution of the sampled geometric objects from the viewpoint of persistent homology. Precisely, we view observed geometric objects through the lens of persistent homology by associating barcodes with each geometric object in the sample. Then we consider the random sample of barcodes associated with random geometric objects to infer the probability distribution of the observed barcodes. Note that the probability distribution of barcodes is well defined (see, e.g., \cite{Mileyko_2011}, \cite{Blumberg2014}). In this article, the goal is to distinguish between two independent collections of random geometric objects up to persistent homology. This amounts to performing a two-sample test for topological signatures of geometric objects computed from persistent homology. 

We consider a two-step hypothesis testing procedure to distinguish two independent collections of random geometric objects. In the first step, we quantify the observed geometric objects using persistence barcodes. Subsequently, in the second step, we formulate a two-sample test for independent collections of random samples of barcodes. The first step towards such testing procedures is to define the class of geometric objects of interest. Therefore, in the following subsection, we first define the geometric objects of interest and then construct an appropriate probability space to incorporate randomness in geometric objects. 

\subsection{Random Geometric Objects}
The class of geometric objects under consideration consists of \emph{compact metric spaces} that are subsets of Euclidean spaces admitting \emph{triangulation}. This class can be defined using the notion of o-minimal structures from \cite{CurryMukherjeeTurner2022} to define tame sets (see Definition \ref{Tame set}) that are triangulable by the Triangulation Theorem in \cite{vandenDries_1998}. In particular, we define the class of geometric objects $\mathcal{X}$ as
\begin{equation} \label{geometric objects}
    \mathcal{X} := \left\{ X \subset \mathbb{R}^d : X \text{ is a tame and compact metric space} \right\}. 
\end{equation}

Next, we define random geometric objects as elements of an appropriate probability space $(\Omega, \mathscr{F}, \mathbb{P})$ that can be constructed as follows. First, one can take $\Omega = \mathcal{X}$ and $\mathscr{F} = \mathcal{B}(\mathcal{X})$, where $\mathcal{B}(\mathcal{X})$ denotes the Borel $\sigma$-algebra generated by the topology induced by a suitable metric on $\mathcal{X}$. In this regard, we can use the Gromov–Hausdorff distance (see Definition \ref{Gromov–Hausdorff distance.}) to define a metric on $\mathcal{X}$. However, the Gromov–Hausdorff distance is a pseudo-metric on $\mathcal{X}$, and a metric on the set of isometry classes of geometric objects is in $\mathcal{X}$ (see \cite{RandomGeometricObjectAOS2465}). Therefore, we define $\Omega$ as:
\begin{equation}\label{Isometery class}
    \Omega := \left\{ [X]: X \in \mathcal{X}\right\},
\end{equation}
where $[X]$ denotes the isometry class of $X$.  

Thus, using the Gromov–Hausdorff metric on $\Omega$, we define the sample space $(\Omega, \mathscr{F})$ on which a suitable probability measure $\mathbb{P}$ can be defined. Note that $\mathscr{F}$ is the Borel $\sigma$-algebra generated by the topology induced by the Gromov–Hausdorff metric on $\Omega$. In what follows, we view random geometric objects as elements from a probability space $(\Omega, \mathscr{F}, \mathbb{P})$, where $\mathbb{P}$ is unknown. 
 


\subsection{Literature Review}
Statistical inference on random geometric objects proceeds with the probability distribution of some suitable geometric summary. In classical statistical shape analysis, geometric objects are represented by a set of user-specified points, known as \emph{landmarks} (\cite{Kendall1989}). The specification of landmarks requires domain knowledge and is subject to bias (\cite{LandmarkBias}). Moreover, landmarks are not suitable for comparing geometric objects, as each geometric object must be represented by an equal number of landmarks (\cite{GaoLandmarking}). \cite{Grenander1998} proposes an alternative to landmark-based approaches to compare geometric objects. However, the approach proposed by \cite{Grenander1998} relies on the assumption that the geometric objects under comparison are diffeomorphic, which may not hold in practice for many data sets.

TDA provides summaries of geometric objects that can be used to distinguish between two collections of geometric objects without specifying landmarks or relying on the assumption that geometric objects are diffeomorphic. In this direction, the following two approaches are relevant. Recently, \cite{RandomShapes2025} proposed a two-way ANOVA testing procedure in the functional data analysis framework by representing geometric objects using the smooth Euler characteristic transform (\cite{Crawford02072020}). In the context of a time series of random geometric objects, two-sample location tests have been proposed in \cite{RandomGeometricObjectAOS2465} for a suitable class of geometric summaries, including dendrograms and persistence diagrams.    

\subsection{Our Contribution}
We introduce a statistical framework for conducting statistical inference on random geometric objects using embeddings from tropical geometry. The proposed framework involves two key steps. First, we quantify geometric objects using persistence barcodes, a topological signature from TDA. Second, we embed the barcodes into a finite-dimensional subset of Euclidean space using embeddings from tropical geometry. The proposed framework differs from the conventional statistical shape analysis framework, which is usually based on manual landmarking of shapes or on the assumption of diffeomorphism of shapes. Thus, the proposed framework paves the way for statistical inference on random geometric objects without relying on landmark- or diffeomorphism-based approaches. Moreover, the proposed framework facilitates statistical analysis of a metric space valued data in an standard statistical framework in a subset of Euclidean space.     

We here present a two-sample test to detect topological and geometric differences in two populations of random geometric objects. A two-step testing procedure is adopted. In the first step, we associate barcodes with the observed random geometric objects, yielding a two-sample problem on the space of barcodes. In the second step, we embed the barcodes in a subset of Euclidean space as statistical analysis on the space of barcodes is prohibited for the reasons explained in the next subsection. To facilitate the testing procedure, a sufficient statistic for persistence barcodes is proposed. The proposed sufficient statistic is based on the tropical embeddings proposed by \cite{Kališnik2019} and refined by \cite{Monod2019} to embed the barcodes in a finite-dimensional Euclidean space. The proposed sufficient statistic complements the main results of \cite{Monod2019} regarding the sufficiency of tropical embeddings by allowing statistical inference for a wider class of probability distributions (see Section \ref{Equivalent Hypothesis Formulation}). Thus, the application of the proposed sufficient statistic translates the two-sample problem on the barcode space into a sample problem on an ordered convex cone $\mathcal{C}_d$ in $\mathbb{R}^d$. We establish that performing a two-sample problem on the space of barcodes is equivalent to performing a two-sample problem on $\mathcal{C}_d$. Finally, we propose a test based on the manifold energy distance to perform a two-sample testing on the manifold $\mathcal{C}_d$. Moreover, the consistency of the proposed test is established. The proposed test can be useful in geometric morphometrics to identify morphological variations in two groups of shapes and for such data analysis.

\subsection{Mathematical Challenges}
We address the following main challenges. We present two-sample tests on the space of barcodes, which enables us to distinguish two independent collections of random geometric objects up to persistent homology. In the testing framework, barcodes are associated with each observed random geometric object, yielding two independent collections of random barcodes. Then, we test the hypothesis of equality of the two probability distributions that generate the two independent collections of random barcodes. However, performing the aforementioned test is prohibited due to the unusual nature of barcodes. Barcodes are collections of intervals in real lines rather than numeric quantities, which makes conventional mathematical operations such as addition and multiplication unavailable to the aforementioned testing framework. 

We circumvent this issue by embedding the barcodes using functions from tropical geometry. In this regard, we use the statistical sufficiency (see Theorem \ref{Sufficieny Result}) of the tropical embeddings proposed by \cite{Monod2019} to embed the barcodes in a finite-dimensional Euclidean space. However, these tropical embeddings are not suitable for two-sample tests as it require stronger assumptions on the distributions of tropical embeddings. In particular, tropical embeddings are sufficient statistics for barcodes if the class of distributions of tropical embeddings is restricted to the class of exchangeable distributions on Euclidean spaces (see Section \ref{Equivalent Hypothesis Formulation}). Therefore, we propose a sufficient statistic using these tropical embeddings that allows us to perform two-sample tests under the standard assumptions on the class of distributions of tropical embeddings. Further, this allows us to establish the equivalence of two-sample problem on the barcode space to the two-sample problem on the the ordered convex cone $ \mathcal{C}_d:= \{(x_1, \ldots, x_d) \in \mathbb{R}^d: x_1\leq, \ldots, \leq x_d\}$. Thus, we can perform two-sample tests on $\mathcal{C}_d$ with standard assumptions on the class of distributions of the proposed sufficient statistic. Furthermore, to perform a two-sample test for manifold-valued data, we propose a permutation test based on manifold energy statistics (\cite{10.1214/23-EJS2203}) and establish its consistency.


\subsection{Organization}
 The rest of the article is organized as follows. Section \ref{Preliminaries} provides the necessary background on the concepts required for the development of the results in the article. Section \ref{Problem Formulation} formulates the hypothesis testing problem of interest and establishes its equivalence to a two-sample problem on a subset of the Euclidean spaces. Section \ref{Test Statistic} presents a testing procedure based on the energy statistics and establish its consistency. Some conclusive remarks are provided in Section \ref{Conclusion}. Finally, technical details such as definitions and proofs of the main results are provided in the appendix in Section \ref{Appendix}.     

\section{Preliminaries} \label{Preliminaries}
This section formalizes the necessary concepts used in the paper, such as the space of barcodes, tropical functions, and tropical coordinates on the space of barcodes. We refer to \cite{kumar2025} and \cite{Kumar2026} for a concise treatment of other required concepts, such as simplicial complexes, simplicial homology, and persistent homology, which are required for this paper.   

\subsection{Barcode Space}
Persistent homology is an adaptation of homology for the filtration of topological spaces. A filtration (right continuous) of topological spaces is a nested collection $\mathcal{F}:= \{\mathcal{F}_{\epsilon}: \epsilon \geq 0\} \text{ such that } \mathcal{F}_{\epsilon}\subseteq \mathcal{F}_{t} \text{ and } \mathcal{F}_{\epsilon} = \cap_{\epsilon < t } \mathcal{F}_{t} \text{ for any } 0 \leq \epsilon \leq t$. In this paper, we consider $\mathcal{F}$ to be a filtration of simplicial complexes. Let $H_k(\mathcal{F}_{\epsilon})$ denote the $k$-dimensional homology of the simplicial complex $\mathcal{F}_{\epsilon} \in \mathcal{F}$, where $k$ is a non-negative integer. We assume that homology groups are computed over field coefficients so that $H_k(\mathcal{F}_{\epsilon})$ is a vector space. Then, the $k$-dimensional persistent homology of $\mathcal{F}$, denoted by $PH_k(\mathcal{F})$, is the indexed family of vector spaces $\{H_k(\mathcal{F}_{\epsilon})\}_{\epsilon \geq 0}$ together with the linear maps $\{\varphi_\epsilon^t: H_k(\mathcal{F}_{\epsilon}) \xrightarrow{} H_k(\mathcal{F}_t)\}_{\epsilon \leq t }$ induced by inclusion $\mathcal{F}_\epsilon \hookrightarrow \mathcal{F}_t$ such that for every $s\leq \epsilon \leq t$, $\varphi_s^t =  \varphi_\epsilon^t \circ \varphi_s^\epsilon $, and $\varphi_t^t$ is the identity map on $H_k(\mathcal{F}_t)$. This family of homology vector spaces, together with linear maps, forms an algebraic structure known as a persistence module (see, e.g., \cite{Chazal2011}). 

In this paper, we consider filtrations obtained by Morse functions (see, e.g., \cite{Milnor1963MorseTheory}) on compact subsets of Euclidean spaces. Hence, the persistence modules under consideration in this paper are tame (see, e.g., \cite{Bubenik2010}). The structure theorem of persistence modules (\cite{ZomorodianCarlsson}) implies that every tame persistence module is finitely generated by interval modules. These interval modules contain isomorphism classes of homological features that appear and disappear through filtration. Essentially, the structure theorem implies that the isomorphism classes of tame persistence modules are in one-to-one correspondence with finite subsets (with multiplicity) of the set $\{[b, d) \in \mathbb{R}^2: 0 \leq b < d < \infty\}$ (see, e.g., \cite{Carlsson(2014)}). Thus, a tame persistence module yields a finite multiset of intervals $\{[b_i, d_i): 0 \leq b_i < d_i < \infty, i \in \mathcal{I}  \}$, known as \emph{persistence barcodes}, where $\mathcal{I}$ is any finite index set. A variant of persistence barcodes is \emph{persistence diagrams} which is the collection $\{(b_i, d_i) \in \mathbb{R}^2: 0 \leq b_i < d_i < \infty, i \in \mathcal{I}  \}$.   

In practice, persistence barcodes are a useful numerical summary of persistent homology. More precisely, for any non-negative integer $k$, a barcode $\mathscr{B}_{k}$ encodes the evolution of k-dimensional homological features (k-cycles) in the filtration of topological spaces by recording the birth and death times of k-dimensional cycles across multiple values of the filtration index $\epsilon$, also called feature scales. The birth time of a k-cycle is the value of $\epsilon$ for which it appears for the first time in the filtration, and the death time refers to the value of $\epsilon$ for which it becomes trivial or merges with any existing k-cycle. In what follows, persistence barcodes are simply referred to as \emph{barcodes}.  

Let $n \in \mathbb{N}$, and consider a barcode $\mathscr{B} := \{[b_{1}, d_{1}), \ldots, [b_{n}, d_{n})\}$, where $b_{i}$ and $d_{i}$ denote the birth time and the death time, respectively, of the $i$-th homological feature. The space of barcodes is the collection of all barcodes, including the barcodes in the set $\Delta = \{ [b, b): b\geq 0\}$, where each point on $\Delta$ has infinite multiplicity. Note that the bars of infinite length are not included in the space of barcodes; however, bars of zero length are allowed with infinite multiplicity. Overall, in view of its complex structure, the space of barcodes is not appropriate for statistical considerations due to its complex geometry. To overcome this issue, various attempts have been made to transform the space of barcodes into some suitable space that admits a vector space structure and has nice geometrical properties without loss of information. For instance, \cite{ADCOCK2016} provides one such approach toward the vectorization of barcodes, which identifies an algebra of polynomials on the space of barcodes. Another approach is due to \cite{Kališnik2019}, which coordinatizes the space of barcodes via tropical symmetric functions. 

The main idea in \cite{ADCOCK2016} and \cite{Kališnik2019} to characterize the space of barcodes is as follows. A barcode with $n$ intervals $\{[b_{1}, d_{1}), \ldots, [b_{n}, d_{n})\}$ is represented as $\{[b_1, \ell_1), \ldots, [b_n, \ell_n) \}$, where $\ell_i = d_i - b_i$ denotes the persistence (or lifetime) of the $i$-th feature for all $i = 1, \ldots, n$. This is the representation of a barcode that we adopt in the rest of the paper. Then, a barcode with $n$ intervals $\{[b_1, \ell_1), \ldots, [b_n, \ell_n) \}$ can be characterized as a vector $(b_1, \ell_1, \ldots, b_n, \ell_n )^\top \in \mathbb{R}^{2n}$. Nevertheless, this characterization is many-to-one, as the barcode structure is invariant to the ordering of intervals $[b_i, \ell_i), i = 1, \ldots, n$. To circumvent this issue, consider the orbit space of the action of the symmetric group on $n$ letters on the product $\left([0, \infty) \times [0, \infty)\right)^n$, given by permuting the coordinates. Further, the goal is to build a natural construction of the space of barcodes of varying cardinality, which informally has the property that intervals of length zero are “ignored”. This construction is exactly the quotient of the set of all barcodes by the equivalence relation that identifies two barcodes that differ only by the addition or deletion of intervals of length zero (see, \cite{ADCOCK2016}). In this paper, we define the space of barcodes in a fixed homological dimension, where each barcode contains finitely many bars. Thus, the space of barcodes consisting of barcodes with at most $n$ intervals, where $n$ is a finite positive integer, is characterized as follows.   

Consider a barcode $\{[b_{1}, \ell_{1}), \ldots, [b_{N}, \ell_{N}) \}$ in a fixed dimension of persistent homology and its associated persistence diagram $\{(b_{1}, \ell_{1}), \ldots, (b_{N}, \ell_{N}) \}$, $N \in [n]$, where $[n] = \{1, \ldots, n\}$. Let $B_{N}$ denotes the orbit space of the symmetric group $S_{N}$ acting on $\{(b_{1}, \ell_{1}), \ldots, (b_{N}, \ell_{N}) \}$. Then the space of barcodes consisting barcodes of at most $n$ intervals denoted by $\mathscr{B}_{\leq n}$ and defined as:
\begin{equation} \label{Barcode space}
    \mathscr{B}_{\leq n} :=  \underset{N \in [n]}{\coprod} B_{N} / \sim,
\end{equation}
where the equivalence relation $\sim$ is defined as 
$$\{[b_{1}, \ell_{1}), \ldots, [b_{N}, \ell_{N}) \} \sim \{[b_{1}, \ell_{1}), \ldots, [b_{M}, \ell_{M}) \} , \text{ whenever } \ell_{N + 1} = \cdots = \ell_{M} = 0, N \leq M \in [n] .$$
In this paper, we represent a barcode $\{[b_{1}, \ell_{1}), \ldots, [b_{n}, \ell_{n}) \}$ by $[(b_{1}, \ell_{1}, \ldots, b_{n}, \ell_{n})] \in \mathscr{B}_{\leq n} $. In what follows, we interchangeably use the term the \emph{barcode space} and the space of barcodes for $\mathscr{B}_{\leq n}$.  

\subsection{ Tropical Functions}
This subsection defines the tropical functions, which serve as the building blocks of the tropical coordinates used in this paper. We refer to \cite{CARLSSON20163610},  \cite{Kališnik2019} for a concise treatment of concepts reviewed in this subsection.   

\begin{d1} \label{tropical def}
The tropical (or, min-plus) semiring is $\left(\mathbb{R}\cup\{+\infty\}, \oplus, \odot \right)$, with the tropical sum $\oplus$ and the tropical multiplication $\odot$ being defined as follows:
    $$ x \oplus y := \text{ min } ( x, y) \text{ and } x \odot y := x + y.$$
Similarly, there exists the max-plus semiring, $\left(\mathbb{R}\cup\{-\infty\}, \boxplus, \odot \right)$, where the multiplication $\odot$ is defined as in the tropical semiring, and the addition $\boxplus$ is defined as $x \boxplus y := \text{ max }(x , y).$ 
\end{d1}
The operators in the min-plus and max-plus semiring have the following properties:
\begin{itemize}
    \item The times operator $\odot$ takes precedence when plus and times occur in the same expression. 

    \item All three operations are commutative and associative.

    \item The distributive law holds:
    $$ x \odot (y \oplus z) = x \odot y \oplus x \odot z \text{ and } x \odot (y \boxplus z) = x \odot y \boxplus x \odot z.$$
\end{itemize}

Let $x_{1}, \ldots, x_{n}$ be elements in the max-plus semiring, then a max-plus monomial expression is defined as the product of these variables, where repetition is allowed. Following is an example of a max-plus monomial expression in three variables $x_{1}, x_{2} \text{ and } x_{3}$:
$$ x_{1} \odot x_{2} \odot x_{3} \odot x_{1} \odot x_{2} = x_{1}^2 \odot x_{2}^2 \odot x_{3}.$$

\begin{d1} \label{tropical poly}
A max-plus polynomial expression, denoted as $p(x_{1},\ldots, x_{n})$, is defined to be a finite linear combination of max-plus monomial expressions. That is, for real numbers $\alpha_{1}, \ldots, \alpha_{m} $, $m \in \mathbb{N}$, we have:
\begin{align*}
    p(x_{1},\ldots, x_{n}) & = \alpha_{1} \odot x_{1}^{r^1_1} x_{2}^{r^1_2} \cdots x_{n}^{r^1_n} \boxplus \alpha_{2} \odot x_{1}^{r^2_1} x_{2}^{r^2_2} \cdots x_{n}^{r^2_n} \boxplus \cdots \boxplus \alpha_{m} \odot x_{1}^{r^m_1} x_{2}^{r^m_2} \cdots x_{n}^{r^m_n}\\
    &= \text{ max } \left( \alpha_{1} + r^1_1 x_{1} + \cdots + r^1_n x_{n}, \ldots, \alpha_{m} + r^m_1 x_{1} + \cdots + r^m_n x_{n}  \right),
\end{align*}
where the exponents $r^i_j$ are non-negative integers for $i = 1, \ldots, m$ and $j = 1, \ldots, n$. The total degree of $p(x_{1},\ldots, x_{n})$ is max $\{r^i_1 + \cdots + r^i_n : i = 1, \ldots, m\}$.     
\end{d1}

Note that each max-plus polynomial expression in $n$ variables can be considered as a function $p : \mathbb{R} ^n \xrightarrow{} \mathbb{R} $ which is continuous, convex, and piecewise linear with a finite number of pieces (\cite{Kališnik2019}). However, max-plus polynomial expressions do not identify functions uniquely; that is, there may exist more than one max-plus polynomial expression that identifies the same function (see, e.g, \cite{Kališnik2019}). This suggests defining an equivalence relation on the class of max-plus polynomial expressions. 

\begin{d1} \label{equivalence}
Let $p$ and $q$ denote two max-plus polynomial expressions in $n$ variables $x_{1}, \ldots, x_{n}$, then $p$ and $q$ are said to be functionally equivalent $i.e$ $p \sim q$ if
$$ p(x_{1}, \ldots, x_{n} ) = q(x_{1}, \ldots, x_{n}), \text{ for all } (x_{1}, \ldots, x_{n}) \in \left(\mathbb{R} \cup \{-\infty\}\right)^n.$$
\end{d1}

Since we are mainly concerned with the functions, instead of observing the entire semiring of max-plus polynomial expressions, we identify those expressions that define the same functions.     

\begin{d1} \label{max-plus poly}
Let $x_{1}, \ldots, x_{n}$ be $n$ variables in the max-plus semiring. Then max-plus polynomials are defined as the semiring of equivalence classes of max-plus polynomial expressions for the equivalence relation defined in the Definition \ref{equivalence}. This semiring is denoted by MaxPlus$\left[x_{1}, \ldots, x_{n}\right]$. We refer functions in MaxPlus$\left[x_{1}, \ldots, x_{n}\right]$ as max-plus functions.       
\end{d1}

In what follows, we are concerned with the max-plus functions in $2n, n \in \mathbb{N}$, variables divided into blocks of two variables each. We denote the semiring of max-plus functions in $2n$ variables as MaxPlus$\left[x_{1, 1}, x_{1, 2} \ldots, x_{n, 1}, x_{n, 2} \right]$. Note that we wish to identify max-plus functions in this semiring defined on the barcode space such that these functions are invariant to the ordering of bars in a barcode. In other words, we want to identify max-plus functions in MaxPlus$\left[x_{1, 1}, x_{1, 2} \ldots, x_{n, 1}, x_{n, 2} \right]$ that are 2-symmetric functions (\cite{CARLSSON20163610}).   
\begin{d1} Let $\psi$ be a max-plus function in $2n$ variables divided into $n$ blocks of 2 variables each. That is, we have $2n$ variables $(x_{1,1},x_{1,2} \ldots, x_{n, 1}, x_{n, 2}) \mapsto \psi(x_{1,1},x_{1,2} \ldots, x_{n, 1}, x_{n, 2})$, then the function $\psi$ is 2-symmetric if it is invariant under the action of symmetric group $S_{n}$ that permutes the blocks of 2-variables $x_{i,1}, x_{i,2}, 1 \leq i \leq n$, while preserving the order of the variables within each block. That is, we have the following:
$$ \psi(x_{1,1},x_{1,2} \ldots, x_{n, 1}, x_{n, 2}) = \psi(x_{\pi(1), 1}, x_{\pi(1), 2} \ldots, x_{\pi(n), 1}, x_{\pi(n), 2} ), \text{ for all }\pi \in S_{n}.$$
\end{d1}

The barcode space is coordinatized using functions in the semiring of 2-symmetric max-plus polynomials (\cite{Kališnik2019}). The semiring of 2-symmetric max-plus polynomials generated using elementary 2-symmetric max-plus polynomials. The elementary 2-symmetric max-plus polynomials are defined as follows. Fix a non-negative integer $n$, and let the symmetric group $S_{n}$ act on the following matrix of indeterminates 
$$X = \begin{pmatrix}
x_{1,1} & x_{1,2}  \\
x_{2,1} & x_{2,2}  \\
\vdots & \vdots  \\
x_{n,1} & x_{n,2}  \\
\end{pmatrix}$$ by left multiplication. Then a max-plus monomial can be determined using the following collection of exponent matrices:
\begin{equation} \label{Exponent Matrices}
    \mathscr{A}_{n} = \left\{ A = \begin{pmatrix}
a_{1,1} & a_{1,2}  \\
a_{2,1} & a_{2,2}  \\
\vdots & \vdots  \\
a_{n,1} & a_{n,2}  \\
\end{pmatrix} \neq \boldsymbol{0}_{n\times 2} : a_{i,j} \in \{0,1\}, i = 1, \ldots n, j = 1, 2.\right\},
\end{equation}
where $\boldsymbol{0}_{n\times 2}$ denotes a null matrix of dimension $n\times 2$. Each matrix $A \in \mathscr{A}_{n}$ determines a max-plus monomial $\rho(A)$ defined as:
$$ \rho(A) = x_{1,1}^{a_{1,1}} x_{1,2}^{a_{1,2}} \cdots x_{n,1}^{a_{n,1}} x_{n,2}^{a_{n,2}}.$$

Let $\mathscr{A}_{n}/ S_{n} $ denote the set of orbits under the row permutation action on $\mathscr{A}_{n}$ by the symmetric group $S_{n}$. That is, we have: 
\begin{equation} \label{orbits}
    \mathscr{A}_{n} /S_{n} = \left\{ \left[\begin{pmatrix}
a_{1,1} & a_{1,2}  \\
a_{2,1} & a_{2,2}  \\
\vdots & \vdots  \\
a_{n,1} & a_{n,2}  \\
\end{pmatrix} \right] : a_{i,j} \in \{0,1\}, i = 1, \ldots n, j = 1, 2.\right\},
\end{equation}
where 
$$  \left[\begin{pmatrix}
a_{1,1} & a_{1,2}  \\
a_{2,1} & a_{2,2}  \\
\vdots & \vdots  \\
a_{n,1} & a_{n,2}  \\
\end{pmatrix} \right] \text{ is an orbit which is a set of matrices obtained by permuting the rows of }  \begin{pmatrix}
a_{1,1} & a_{1,2}  \\
a_{2,1} & a_{2,2}  \\
\vdots & \vdots  \\
a_{n,1} & a_{n,2}  \\
\end{pmatrix} .$$

Then, for any orbit $\Lambda \in \mathscr{A}_{n} /S_{n} $, an elementary 2-symmetric max-plus polynomial can be determined as
\begin{equation} \label{Elementary maxplus}
    \rho(A_{1}) \boxplus \cdots \boxplus \rho(A_{m}),
\end{equation}
where $ m = n! \text{ , and } A_{1}, \ldots, A_{m}$ are matrices of dimension $n \times 2$ in the orbit  $\Lambda$. We simplify notation by writing $[(a_{1, 1}, a_{1,2}), \ldots, (a_{n, 1}, a_{n,2}) ]$ for any orbit $\Lambda \in \mathscr{A}_{n} /S_{n}$ such that 
$$\Lambda  = \left[\begin{pmatrix}
a_{1,1} & a_{1,2}  \\
a_{2,1} & a_{2,2}  \\
\vdots & \vdots  \\
a_{n,1} & a_{n,2}  \\
\end{pmatrix}\right],$$ 
then $\Gamma_{\displaystyle{[(a_{1,1}, a_{1,2}), \ldots, (a_{n,1}, a_{n,2})]} }$ denotes a 2-symmetric max-plus polynomial that arise from the orbit $\Lambda$.   

\begin{e1} \label{First Example} A barcode with $n = 2$ intervals can be written as $ \mathcal{B} :=\{[b_{1}, \ell_{1}), [b_{2}, \ell_{2})\} \in \mathscr{B}_{\leq 2}$. The set of orbits under the action of the symmetric group $S_{2}$ on $\mathscr{A}_{2}$ (see, Equation \eqref{Exponent Matrices}) is given by

\begin{equation} \label{set of orbits}
    \mathscr{A}_{2} / S_{2} =  \left\{\begin{array}{ccccc}
 \left[\begin{pmatrix} 1 & 1  \\ 1 & 0  \\ \end{pmatrix}\right], \left[\begin{pmatrix} 0 & 1  \\ 1 & 1  \\ \end{pmatrix}\right],\left[\begin{pmatrix} 0 & 0  \\ 1 & 1  \\ \end{pmatrix}\right], \left[\begin{pmatrix} 1 & 0  \\ 1 & 0  \\ \end{pmatrix}\right],\\ \\ \left[\begin{pmatrix} 0 & 1  \\ 1 & 0 \\ \end{pmatrix}\right], \left[\begin{pmatrix} 0 & 1  \\ 0 & 1  \\ \end{pmatrix}\right], \left[\begin{pmatrix}  0 & 0  \\ 0 & 1  \\ \end{pmatrix}\right], \left[\begin{pmatrix} 0 & 0  \\ 1 & 0  \\ \end{pmatrix}\right], \left[\begin{pmatrix} 1 & 1  \\ 1 & 1  \\ \end{pmatrix}\right]\end{array}
 \right\} .
\end{equation}

Then the following elementary 2-symmetric max-plus polynomial denotes the maximum persistence of topological features denoted by the two bars in $\mathcal{B}$.
\begin{align*}
    \Gamma_{\displaystyle{[(0, 1) , (0, 0)]}}\displaystyle{\left[\left( b_{1}, \ell_{1}, b_{2}, \ell_{2} \right)\right]} &= \rho(A_{1}) \boxplus \rho(A_{2}) \\ 
    &= b_{1}^0 \odot \ell_{1}^1 \odot b_{2}^0 \odot \ell_{2}^0 \boxplus b_{1}^0 \odot \ell_{1}^0 \odot b_{2}^0 \odot \ell_{2}^1 \\
    &= \ell_{1} \boxplus \ell_{2}\\
    &= \max(\ell_{1}, \ell_{2}),
    \end{align*}
    where $$ [(0, 1) , (0, 0)]\text{ denotes the orbit } \left[ \begin{pmatrix} 0 & 1 \\ 0 & 0
        
    \end{pmatrix}\right] \text{ consisting the matrices } A_{1} = \begin{pmatrix}  0 & 1 \\ 0 & 0 \end{pmatrix} \text{ and } A_{2} = \begin{pmatrix} 0 & 0 \\ 0 & 1 \end{pmatrix}.$$

Similarly, a few more examples of 2-symmetric max-plus polynomials are as follows:  
$$ \Gamma_{\displaystyle{[(1, 0), (0, 0)]}}\displaystyle{[(b_{1}, \ell_{1}, b_{2}, \ell_{2} )]} = b_{1} \boxplus b_{2} = \max (b_{1}, b_{2}),$$
$$\Gamma_{\displaystyle{[(1, 1), (1, 1)]}}[(b_{1}, \ell_{1}, b_{2}, \ell_{2} )] = b_{1} \odot \ell_{1} \odot b_{2} \odot \ell_{2} = b_{1} + \ell_{1} + b_{2} + \ell_{2} ,$$
$$\Gamma_{\displaystyle{[(1, 1), (0, 1)]}}[(b_{1}, \ell_{1}, b_{2}, \ell_{2} )] = b_{1} \odot \ell_{1} \odot \ell_{2} \boxplus \ell_{1} \odot b_{2} \odot  \ell_{2}   = \max (b_{1} + \ell_{1} + \ell_{2}, \ell_{1} + b_{2} + \ell_{2}) .$$
\end{e1}

We simplify notation by $\displaystyle{\Gamma}_{\displaystyle{[(0, 1)]}}$ instead of $\Gamma_{\displaystyle{[(0, 1) , (0, 0)]}}$, when $n$ is clear from the context. Similarly, we write $\Gamma_{\displaystyle{[(1, 1)^2]}}$ instead of $\Gamma_{\displaystyle{[(1, 1) , (1, 1)]}}$. This is the notational convention that we follow in the rest of the paper. In general, when $n$ is known, any elementary 2-symmetric max-plus polynomial $\Gamma_{\displaystyle{[(a_{1,1}, a_{1,2}), \ldots, (a_{n,1}, a_{n,2})]} }$ is denoted by $\Gamma_{\displaystyle{[(0, 1)^{i}, (1, 1)^{j}, (1, 0)^{k}]}}$, where $i$ denotes the number of (0, 1) rows, $j$ denotes the number of (1, 1) rows, $k$ denotes the number of (1, 0) rows in the matrix having $n$ rows $(a_{1,1}, a_{1,2}), \ldots, (a_{n,1}, a_{n,2})$ that constitutes the orbit $\left[(a_{1,1}, a_{1,2}), \ldots, (a_{n,1}, a_{n,2})\right] \in \mathscr{A}_{n} /S_{n}$ (see, Equation \eqref{orbits}). The number of (0, 0) rows is what remains, hence it is omitted from the notation. In other words, the number of (0, 0) rows is $n - i - j - k$. In this paper, the functions $\Gamma_{\displaystyle{[(0, 1)^{i}, (1, 1)^{j}, (1, 0)^{k}]}}$ for $i, j, k \in \{0, \ldots,n\}$ such that $(i + j + k) \leq n$, are referred to as \emph{tropical functions}.

\subsection{ Tropical Embedding } \label{Coordinates on the barcode space}
This subsection defines embeddings of barcodes into a finite-dimensional Euclidean space using tropical functions defined in the previous section. In this context, the following result from \cite{Kališnik2019} will be useful.  
\begin{t1} \label{Tropical coordinates} (Theorem 6.3 of \cite{Kališnik2019}) Fix $n \in \mathbb{N}$, and for $i, j, k \in \{0, \ldots,n\}$ such that $(i + j + k) \leq n$, consider the family of functions $\left\{\displaystyle{T_m^{(i,j,k)} : m \in \mathbb{N}}\right\}$ on $\mathscr{B}_{\leq n}$, defined by
$$ T_m^{(i,j,k)}([(b_{1}, \ell_{1}, \ldots, b_{n}, \ell_{n} )]) := \Gamma_{\displaystyle{[(0, 1)^{i}, (1, 1)^{j}, (1, 0)^{k}]}} [( b_{1} \oplus \ell_{1}^m, \ell_{1}, \ldots, b_{n} \oplus \ell_{n}^m, \ell_{n})].$$
Then, for two distinct point $\mathscr{B}_1$ and $\mathscr{B}_2$ in $\mathscr{B}_{\leq n}$, there exists $(i, j, k)$ such that $T_m^{(i,j,k)}(\mathscr{B}_1) \neq T_m^{(i,j,k)}(\mathscr{B}_2) $, for $m \in \mathbb{N}$. In other words, the family of functions $\left\{\displaystyle{T_m^{(i,j,k)} : m \in \mathbb{N}}\right\}$ separate nonequivalent points in $\mathscr{B}_{\leq n}$.  
\end{t1}

 In essence, Theorem \ref{Tropical coordinates} provides an infinite list of real-valued functions that can be used as coordinates on $\mathscr{B}_{\leq n}$. However, working with infinite-dimensional vectors is not feasible in practice. Therefore, as a viable solution to this problem, \cite{Monod2019} suggests working with a regularized barcode space for a fixed $m \in \mathbb{N}$. The \emph{regularized barcode space} is denoted by $\mathscr{B}_{\leq n}^m$ and defined as:
 \begin{equation} \label{regularized barcode space}
    \mathscr{B}_{\leq n}^m := \left\{ \left[(b_1, \ell_1, \ldots, b_n, \ell_n)\right] \in \mathscr{B}_{\leq n} : b_i \leq m \ell_i, i = 1, \ldots, s \right\},
\end{equation} 
where $s = \sum_{i = 1}^n\mathds{1}(\ell_i > 0)$, and $\mathds{1}$ denote the indicator function.

Note that restricting $\mathscr{B}_{\leq n}$ to $\mathscr{B}_{\leq n}^m$ does not pose any limitations in practice, as important properties such as Lipschitz continuity of the functions defined in Theorem \ref{Tropical coordinates} that hold on $\mathscr{B}_{\leq n}$, also remain intact on $\mathscr{B}_{\leq n}^m$. The advantage of working with $\mathscr{B}_{\leq n}^m$ instead of $\mathscr{B}_{\leq n}$, as the barcode space, is that the functions defined in Theorem \ref{Tropical coordinates} provide an embedding of barcodes into a finite-dimensional Euclidean space. Moreover, for a given finite set of barcodes, a choice of $m$ is straightforward. For example, we can take $m = \lceil \underset{1 \leq i \leq s}{\max}(b_{i}/\ell_{i}) \rceil $, where $s = \sum_{i = 1}^n\mathds{1}(\ell_i > 0) $, and $\lceil x \rceil$ denotes the smallest integer greater than or equal to $x$ for any $x \in \mathbb{R}$.  

Therefore, \cite{Monod2019} presented a modified version of Theorem \ref{Tropical coordinates} for the regularized barcode space $\mathscr{B}_{\leq n}^m$ and for the tropical functions parameterized by $i, j \in \{0, \ldots,n\} $ alone so that the $k$-factor of (1, 0) rows is redundant. In particular, for a fixed $m \in \mathbb{N}$, consider the family of functions $\{T^{(i, j)}: i, j \in \{0, \ldots,n\} \text{ such that } (i+ j) \leq n \}$ on $\mathscr{B}_{\leq n}^m$, defined by
 \begin{equation} \label{Finite Coordinates}
   T^{(i, j)}([(b_{1}, \ell_{1}, \ldots, b_{n}, \ell_{n} )]) := \Gamma_{\displaystyle{[(0, 1)^{i}, (1, 1)^{j}]}} [( b_{1} \oplus \ell_{1}^m, \ell_{1}, \ldots, b_{n} \oplus \ell_{n}^m, \ell_{n})].  
\end{equation} 
The functions defined in Equation \eqref{Finite Coordinates} separate the nonequivalent barcodes in $\mathscr{B}_{\leq n}^m$ and are Lipschitz with respect to the bottleneck distance (see Definition \ref{Bottleneck distance}). In what follows, we denote the family of functions $\{T^{(i, j)}: i, j \in \{0, \ldots,n\} \text{ such that } (i+ j) \leq n \}$ by $\{T_1, \ldots, T_d\}$, $d$ is determined by equation $2d = 2n + n(n+1)$ (see \cite{Monod2019}), and call the functions $T_1, \ldots, T_d$ as \emph{tropical coordinates} on $\mathscr{B}_{\leq n}^m$.

Thus, given a barcode $\mathscr{B} \in \mathscr{B}_{\leq n}^m$ and tropical coordinates $\{T_1, \ldots, T_d\}$ on $\mathscr{B}_{\leq n}^m$, we have an embedding $\mathcal{T}: \mathscr{B}_{\leq n}^m \xrightarrow{} \mathbb{R}^d$, defined as:
\begin{equation} \label{Tropical Embedding}
    \mathcal{T}\left(\mathscr{B}\right):= \left(T_{\pi(1)}\left(\mathscr{B}\right), \ldots, T_{\pi(d)}\left(\mathscr{B}\right)\right)^\top,
\end{equation}
where $\pi$ is a fixed permutation on $\{1, \ldots,d\},d = n + 0.5n(n+1)$. In general, we have $d!$ embeddings in $\mathbb{R}^d$ for a barcode $\mathscr{B} \in \mathscr{B}_{\leq n}^m$. This fact is crucial for the development of the methodology proposed in this paper. The following example illustrates the computation of tropical coordinates on a regularized barcode space, and thereby an embedding in a Euclidean space. 

\begin{e1} \label{MExample} 
Recall in Example \ref{First Example}, for $n = 2$, the set of orbits under the row permutation action of the symmetric group $S_{2}$ on $\mathscr{A}_{2}$ (see, Equation \eqref{Exponent Matrices}), denoted by $\mathscr{A}_{2}/ S_{2}$, is given in Equation \eqref{set of orbits}. According to Proposition 2.8 of \cite{Monod2019}, it suffices to work with the following subsets of $\mathscr{A}_{2}/ S_{2}$ to define tropical coordinates on the barcode space:
\begin{equation} \label{suborbits}
    \left\{\begin{array}{ccccc}
 \left[\begin{pmatrix} 0 & 1  \\ 1 & 1  \\ \end{pmatrix}\right],\left[\begin{pmatrix} 0 & 0  \\ 1 & 1  \\ \end{pmatrix}\right], \left[\begin{pmatrix} 0 & 1  \\ 0 & 1  \\ \end{pmatrix}\right], \left[\begin{pmatrix}  0 & 0  \\ 0 & 1  \\ \end{pmatrix}\right], \left[\begin{pmatrix} 1 & 1  \\ 1 & 1  \\ \end{pmatrix}\right]\end{array}
 \right\} . 
\end{equation}
 Now, suppose we are given two barcodes in $\mathscr{B}_{\leq 2}$ denoted by $\mathscr{B}_{1}$ and $\mathscr{B}_{2}$, where $\mathscr{B}_{1} = [[2, 1), [3, 1)]$, and $\mathscr{B}_{2} = [[4, 4)]$. In the first step, we compute $m$ for the observed bars. In this example $m = 3$, since $\lceil\underset{ 1 \leq i \leq 3 }{\max}(b_{i} / \ell_{i}) \rceil = \lceil \max (2/1, 3/1, 1) \rceil= 3$. This implies that $\mathscr{B}_{1}, \mathscr{B}_{2} \in \mathscr{B}_{\leq 2}^3$. Note that the functions $T^{(i,j)}$ (see Equation \eqref{Finite Coordinates}) will be obtained from the orbits in Equation \eqref{suborbits}. In particular, for the first orbit in Equation \eqref{suborbits}, that is, for $i = 1$ (number of (0 1) rows) and $j = 1$ (number of (1 1) rows), we will have $T^{(1, 1)}$, which we denote by $T_1$. Similarly, we will have $T_2, T_3, T_4, T_5$ corresponding to the rest of the orbits in Equation \eqref{suborbits}. Let $\mathscr{B} = [(b_{1}, \ell_{1}, b_{2}, \ell_{2}  )] \in \mathscr{B}^3_{\leq 2}$, then the tropical coordinates $T_1, \ldots, T_5$ on $\mathscr{B}^3_{\leq 2}$ are obtained as follows.       
\begin{align*}
     T_1(\mathscr{B}) := T^{(1, 1)} [(b_{1}, \ell_{1}, b_{2}, \ell_{2}  )] &= \Gamma_{\displaystyle{[(0, 1), (1,1)]}} \left( b_{1} \oplus \ell_{1}^3, \ell_{1}, b_{2} \oplus \ell_{2}^3, \ell_{2} \right)\\
     &= \left(\ell_{1} \odot \left(b_{2} \oplus \ell_{2}^3 \right) \odot \ell_{2} \right) \boxplus \left( \left(b_{1} \oplus \ell_{1}^3 \right) \odot \ell_{1} \odot \ell_{2}  \right) \\
     &= \max \left(\ell_{1} \odot \left(b_{2} \oplus \ell_{2}^3\right) \odot \ell_{2} \text{, } \left(b_{1} \oplus \ell_{1}^3 \right) \odot \ell_{1} \odot \ell_{2}\right )\\
     &= \max \left(\ell_{1} \odot \min \left(b_{2} \text{, } \ell_{2}^3\right) \odot \ell_{2} \text{, } \min \left(b_{1} \text{, } \ell_{1}^3 \right) \odot \ell_{1} \odot \ell_{2}\right )\\
     &= \max \left(\ell_{1} + \min \left(b_{2} \text{, } \ell_{2}^3\right) + \ell_{2} \text{, } \min \left(b_{1} \text{, } \ell_{1}^3 \right) + \ell_{1} + \ell_{2}\right )\\
     &= \max \left(\ell_{1} + \min \left(b_{2} \text{, } 3\ell_{2}\right) + \ell_{2} \text{, } \min \left(b_{1} \text{, } 3\ell_{1}\right) + \ell_{1} + \ell_{2}\right ).
\end{align*}
\begin{align*}
    T_2(\mathscr{B}) := T^{(0, 1)} [(b_{1}, \ell_{1}, b_{2}, \ell_{2}  )] & = \Gamma_{\displaystyle{[(1, 1)]}} [(b_{1}, \ell_{1}, b_{2}, \ell_{2}  )]
      = \max \left( \min \left( b_{1}, 3 \ell_{1} \right) + \ell_{1}, \min \left( b_{2}, 3 \ell_{2} \right) + \ell_{2}\right).
\end{align*}
 \begin{align*}
     T_3(\mathscr{B}):= T^{(2, 0)}[(b_{1}, \ell_{1}, b_{2}, \ell_{2}  )] & = \Gamma_{\displaystyle{[(0, 1)^2]}} [(b_{1}, \ell_{1}, b_{2}, \ell_{2}  )]
      = \ell_{1} + \ell_{2}. 
 \end{align*}
 \begin{align*}
     T_4(\mathscr{B}):= T^{(1, 0)} [(b_{1}, \ell_{1}, b_{2}, \ell_{2}  )] & = \Gamma_{\displaystyle{[(0, 1)]}} [(b_{1}, \ell_{1}, b_{2}, \ell_{2}  )]
      = \max (\ell_{1} , \ell_{2}). 
 \end{align*}
 \begin{align*}
     T_5(\mathscr{B}):= T^{(0,2)} [(b_{1}, \ell_{1}, b_{2}, \ell_{2}  )] & = \Gamma_{\displaystyle{[(1, 1)^2]}} [(b_{1}, \ell_{1}, b_{2}, \ell_{2}  )]
      = \min (b_{1}, 3\ell_{1}) + \ell_{1} + \min (b_{2}, 3\ell_{2}) + \ell_{2}. 
 \end{align*}
Thus, we have tropical coordinates $\{T_1,\ldots, T_5\}$ on $\mathscr{B}_{\leq 2}^3$. Consequently, the barcodes $\mathscr{B}_{1} = \{[2,1), [3, 1)\}$ and $\mathscr{B}_{2} = \{ [4, 4)\}$ can be embedded in $\mathbb{R}^5$ as $(5, 4, 2, 1, 7)^T$ and $(8, 8, 4, 4, 8 )^T$, respectively, by taking $\pi(i) = i, i= 1, \ldots, 5$ in Equation \eqref{Tropical Embedding}.  
\end{e1}

\section{Problem Formulation} \label{Problem Formulation}
Let $X_1, \ldots, X_{n_1} \overset{i.i.d}{\sim} \mathbb{P}_1$ and $Y_1, \ldots, Y_{n_2} \overset{i.i.d}{\sim} \mathbb{P}_2$ be two independent random samples of geometric objects, where the probability measures $\mathbb{P}_1$ and $\mathbb{P}_2$ are defined on the measurable space $(\Omega, \mathscr{F})$. Here $n_1$ and $n_2$ denotes the sample sizes, which may or may not be equal. Recall that $\Omega$ is defined in Equation \eqref{Isometery class}, and $\mathscr{F} = \mathcal{B}(\Omega)$, where $\mathcal{B}(\Omega)$ is the Borel $\sigma$-algebra generated by the topology induced by the Gromov–Hausdorff metric (denoted by $d_{GH}$, see Definition \ref{Gromov–Hausdorff distance.}) on $\Omega$. That is, $\mathcal{B}(\Omega)$ is the smallest $\sigma$-algebra that contains all open sets in the metric space $(\Omega, d_{GH})$ with topology generated by open balls. We aim to detect topological differences between two independent collections of random geometric objects. Our approach quantifies each geometric object using topological signatures and then performs a two-sample hypothesis testing for the probability distributions of the topological signatures. In this paper, we quantify the topological content of geometric objects using persistence barcodes.

Let $\gamma: (\Omega, \mathscr{F}) \xrightarrow{} (\mathscr{B}_{\leq n}, \sigma (\mathscr{B}_{\leq n}) )$ be a measurable transformation, where $\sigma (\mathscr{B}_{\leq n})$ is the smallest $\sigma$-algebra that contains all open sets in the metric space $(\mathscr{B}_{\leq n}, \delta_B)$ with topology generated by open balls. Here, $\delta_B$ denotes the bottleneck distance (see Definition \ref{Bottleneck distance}). Then, for a geometric object $X \in \Omega$, $\gamma(X)$ is a topological signature representing the persistence barcode of $X$ and if $X \sim \mathbb{P}$, then $\gamma(X) \sim \mathbb{P}\circ \gamma^{-1}$, $\mathbb{P}\circ \gamma^{-1}$ denotes the push-forward of $\mathbb{P}$ under $\gamma$. We are interested in the following two-sample hypothesis testing problem: 
\begin{equation}\label{primary test}
    \mathcal{H}_0^\prime: \mu(A) = \nu(A) \text{ for all } A \in \sigma (\mathscr{B}_{\leq n}) \text{ vs. } \mathcal{H}_1^\prime: \mu(A) \neq \nu(A), \text{ for some } A \in \sigma (\mathscr{B}_{\leq n}),
\end{equation}
where $\mu \equiv \mathbb{P}_1\circ\gamma^{-1}$ and $\nu \equiv \mathbb{P}_2\circ\gamma^{-1}$.

In essence, the testing problem in Equation \eqref{primary test} is a two-sample problem on the barcode space for the two independent random samples of barcodes $\mathscr{B}_{1}, \ldots, \mathscr{B}_{n_{1}} \overset{i.i.d}{\sim} \mu $ and $\boldsymbol{\Tilde{\mathscr{B}}}_{1}, \ldots, \boldsymbol{\Tilde{\mathscr{B}}}_{n_{2}} \overset{i.i.d}{\sim} \nu$. However, devising a testing procedure using barcodes as data points is prohibitive due to the complex nature of barcodes. In particular, the usual mathematical operations, such as addition and multiplication, cannot be applied to a collection of intervals in $\mathbb{R}$. Therefore, to place the testing framework on a standard statistical footing in Euclidean space, we formulate an equivalent hypothesis on Euclidean space by regularizing the observed barcodes for a suitable $m \in \mathbb{N}$ and then using the tropical embeddings (see Equation \eqref{Tropical Embedding}) on the regularized barcode space $\mathscr{B}_{\leq n}^m$ (see Equation \eqref{regularized barcode space}). 

\subsection{Equivalent Hypothesis Formulation} \label{Equivalent Hypothesis Formulation}
We translate the two-sample problem on the barcode space defined in Equation \eqref{primary test} to a two-sample problem on the Euclidean space using tropical embeddings defined in Equation \eqref{Tropical Embedding}. In this context, the following result from \cite{Monod2019} regarding the statistical sufficiency of tropical embeddings will be useful.
\begin{t1} \label{Sufficieny Result} (Theorem 3.5 of \cite{Monod2019}) Consider a statistical model on $(\mathscr{B}_{\leq n}^m, \sigma(\mathscr{B}_{\leq n}^m))$ with a family of probability measures $\mathcal{P}$ dominated by a $\sigma$-finite measure $\lambda$, then for a barcode $\mathscr{B} \sim \vartheta \in \mathcal{P}$, the embedding $\mathscr{B}\mapsto \mathcal{T}(\mathscr{B}) = (T_{\pi(1)}(\mathscr{B}), \ldots, T_{\pi(d)} (\mathscr{B}))^\top \in \mathbb{R}^d $ (see Equation \eqref{Tropical Embedding}), for a fixed permutation $\pi$ on $\{1, \ldots,d\}$, is a sufficient statistic for $\mathcal{P}$. In other words, for each $\vartheta \in \mathcal{P}$, the Radon-Nikodym derivative $f_{\vartheta} \equiv d\vartheta/d\lambda$ admits the factorization 
$$ f_{\vartheta} (\mathscr{B}) = h(\mathscr{B})g_{\vartheta}(\mathcal{T}(\mathscr{B})),$$
where $h$ is a non-negative measurable function on $\mathscr{B}_{\leq n}^m$, $g_{\vartheta}$ is a non-negative measurable function on $\mathbb{R}^d, d = n + 0.5n(n +1)$, $\sigma(\mathscr{B}_{\leq n}^m)$ denotes the smallest $\sigma$-algebra that contains all open sets in the metric space $(\mathscr{B}_{\leq n}^m, \delta_B)$ with topology generated by open balls, $\delta_B$ denotes the bottleneck distance (see Definition \ref{Bottleneck distance}).  
\end{t1}

In view of Theorem \ref{Sufficieny Result}, we can regularize the barcode space $\mathscr{B}_{\leq n}$ for a suitable choice of $m \in \mathbb{N}$, and consider the two random samples of barcodes $\mathscr{B}_{1}, \ldots, \mathscr{B}_{n_{1}} \overset{i.i.d}{\sim} \mu $ and $\boldsymbol{\Tilde{\mathscr{B}}}_{1}, \ldots, \boldsymbol{\Tilde{\mathscr{B}}}_{n_{2}} \overset{i.i.d}{\sim} \nu$ such that $\mu$ and $\nu$ are defined on $(\mathscr{B}_{\leq n}^m, \sigma(\mathscr{B}_{\leq n}^m))$. Then, Theorem \ref{Sufficieny Result} can be used to formulate an equivalent hypothesis to Equation \eqref{primary test} based on the random samples $\mathcal{T}(\mathscr{B}_{1}), \ldots, \mathcal{T}(\mathscr{B}_{n_{1}}) \overset{i.i.d}{\sim} F$ and $\mathcal{T}(\Tilde{\mathscr{B}_{1}}), \ldots, \mathcal{T}(\Tilde{\mathscr{B}_{n_{2}}}) \overset{i.i.d}{\sim} G$, where $F$ and $G$ are probability distributions on $\mathbb{R}^d$, where $d \geq 2$ for $n \geq 1$, by the equation $2d = 2n + n(n + 1)$. Recall that $n$ denotes the maximum number of features (bars) in a barcode $\mathscr{B} \in\mathscr{B}_{\leq n}$. Note that $F$ and $G$ could be considered as continuous probability distributions, since tropical embeddings are continuous due to the Lipschitz continuity of tropical coordinates with respect to the bottleneck distance (see \cite{Kališnik2019}).    

However, Theorem \ref{Sufficieny Result} is valid under the assumption that the probability distributions $F$ and $G$ belong to the class of exchangeable distributions. This is because for any two tropical embeddings $\mathcal{T}_\pi$ and $\mathcal{T}_\sigma$ corresponding to the two different permutations $\pi$ and $\sigma$ on $\{1, \ldots, d\}$, respectively, Theorem \ref{Sufficieny Result} yields the following by taking $h(\mathscr{B}) \equiv 1$ without loss of generality:  
\begin{equation} \label{Identifiablility}
    f_{\vartheta} (\mathscr{B}) = g_{\vartheta}(\mathcal{T}_\pi(\mathscr{B})) = \Tilde{g}_{\vartheta}(\mathcal{T}_\sigma(\mathscr{B})),
\end{equation}
where $g_{\vartheta}$ and $\Tilde{g}_{\vartheta}$ are the probability densities of $\mathcal{T}_\pi(\mathscr{B}) := (T_{\pi(1)} (\mathscr{B}), \ldots, T_{\pi(d)} (\mathscr{B}) )^\top$ and $\mathcal{T}_\sigma(\mathscr{B}):= (T_{\sigma(1)}(\mathscr{B}), \ldots, T_{\sigma(d)} (\mathscr{B}) )^\top$, respectively. Here,  $\Tilde{g}_{\vartheta} \equiv g_{\vartheta} \circ \phi$, $\phi$ is a bijection such that $\phi (\mathcal{T}_\sigma(\mathscr{B})):= \mathcal{T}_\pi(\mathscr{B})$ and $T_i$'s are tropical coordinates defined in Equation \eqref{Finite Coordinates}. Thus, if the densities $g_{\vartheta}$ and $\Tilde{g}_{\vartheta}$ with respect to the induced probability measures $\vartheta \circ \mathcal{T}_\pi^{-1}$ and $\vartheta \circ \mathcal{T}_\sigma^{-1}$, respectively, are not the same, then for a point $\mathscr{B} \in \mathscr{B}^m_{\leq n}$, there will be two images of $\mathscr{B}$ under $f_{\vartheta}$. This implies that $f_\vartheta$ will not be a map from $\mathscr{B}_{\leq n}^m$ to $(0, \infty)$. Consequently, Theorem \ref{Sufficieny Result} will not be valid as $f_{\vartheta}$ will not be a probability density with respect to the probability measure $\vartheta \in \mathcal{P}$, defined on $(\mathscr{B}_{\leq n}^m, \sigma(\mathscr{B}_{\leq n}^m))$. Hence, the probability distributions with respect to the induced measures $\vartheta \circ \mathcal{T}_\pi^{-1}$ and $\vartheta \circ \mathcal{T}_\sigma^{-1}$ need to be exchangeable for the validity of Theorem \ref{Sufficieny Result}.      

Note that the condition in Equation \eqref{Identifiablility} is trivially true if we assume that the tropical coordinates are independent and identically distributed (i.i.d). However, in the present context, the i.i.d assumption for the tropical coordinates is too restrictive, as an individual component of the tropical coordinates does not represent a barcode in the barcode space. In addition, the class of exchangeable distributions excludes some important classes of distributions, such as $\{\mathcal{N} (\boldsymbol{\theta}_{d \times1}, \Sigma): \boldsymbol{\theta}_{d \times1} \neq \theta\boldsymbol{1}_{d \times 1},  \Sigma \neq \sigma^2\mathcal{I}_{d\times d}, \theta \in \mathbb{R},\sigma^2 > 0, d \geq 2\}$. Moreover, we need to validate the assumption via testing the hypothesis whether the observed vector representations are exchangeable or not.  Therefore, to allow testing framework for a wider class of probability distributions, we propose an embedding based on the tropical coordinates defined in Equation \eqref{Finite Coordinates} and establish that the embedding is a sufficient statistic. In the following theorem, we define the proposed embedding and state its statistical sufficiency. 

\begin{t1} \label{Modified Sufficiency}  Consider a statistical model on $(\mathscr{B}_{\leq n}^m, \sigma(\mathscr{B}_{\leq n}^m))$ with a family of probability measures $\mathcal{P}$ dominated by a $\sigma$-finite measure $\lambda$, then for a barcode $\mathscr{B} \sim \vartheta \in \mathcal{P}$, the map $\mathscr{B}\mapsto \mathcal{V}(\mathscr{B}) := (V_1(\mathscr{B}), \ldots, V_d(\mathscr{B}))^\top \in \mathcal{C}_d $, $V_k:= \text{k-}\min \{T_1(\mathscr{B}), \ldots, T_d(\mathscr{B})\}, k = 1, \ldots, d$, is a sufficient statistic for $\mathcal{P}$, where $\mathcal{C}_d := \{(x_1, \ldots, x_d)^\top \in \mathbb{R}^d: x_1 \leq x_2 \leq \ldots, \leq x_d\}$   and the $\text{k-}\min$ denotes the $k$-th smallest value from the tropical coordinates $\{T_1(\mathscr{B}), \ldots, T_d(\mathscr{B})\}$ (see Equation \eqref{Finite Coordinates}).
\end{t1}
We refer to Section \ref{Proofs} for the proof of Theorem \ref{Modified Sufficiency}. In fact, the embedding $\mathcal{V}$ is a minimal sufficient statistic among all sufficient statistics generated by the tropical embeddings defined in Equation \eqref{Tropical Embedding}. This is because, corresponding to every sufficient statistic generated by tropical embeddings, there exists a measurable function $\Psi$ such that $\mathcal{V}(\mathscr{B}) = \Psi(\mathcal{T}(\mathscr{B}))$, where $\Psi$ is a map that sort the elements in the vector $\mathcal{T}(\mathscr{B})$ in increasing order. Hence, by the definition of a minimal sufficient statistic (see, e.g., Definition 2.5 of \cite{JunShao}), the sufficient statistic $\mathcal{V}$ is a minimal sufficient statistic.

Thus, as an application of Theorem \ref{Modified Sufficiency}, we propose to use the minimal sufficient statistic to formulate an equivalent hypothesis to the hypothesis in Equation \eqref{primary test}. This allows us to perform two-sample tests for the hypothesis in Equation \eqref{primary test} by two-sample tests on the manifold $\mathcal{C}_d$. Note that we do not require probability distributions on $\mathcal{C}_d$ to be exchangeable to perform two-sample tests on the barcode space.         

Now, we present the main result that states that two-sample tests on $\left(\mathscr{B}^m_{\leq n}, \sigma (\mathscr{B}^m_{\leq n})\right)$ can be performed using the probability measures on the manifold $\mathcal{C}_{d}$, using the minimal sufficient statistic from Theorem \ref{Modified Sufficiency}.   

\begin{t1} \label{Equivalence of Hypothesis}
Suppose we observe two independent samples of barcodes $\mathscr{B}_{1}, \ldots, \mathscr{B}_{n_{1}} \overset{i.i.d}{\sim} \mu$, and $\Tilde{\mathscr{B}_{1}}, \ldots, \Tilde{\mathscr{B}_{n_2}} \overset{i.i.d}{\sim} \nu$, where $\mu$ and $\nu$ are probability measures defined on $(\mathscr{B}_{\leq n}^m, \sigma(\mathscr{B}_{\leq n}^m)$. Let the tropical representation of barcodes be $\mathcal{V}(\mathscr{B}_{1}), \ldots, \mathcal{V}(\mathscr{B}_{n_{1}}) \overset{i.i.d}{ \sim} F$ and $\mathcal{V}(\Tilde{\mathscr{B}_{1}}), \ldots, \mathcal{V}(\Tilde{\mathscr{B}_{n_{2}}}) \overset{i.i.d}{ \sim} G$, where $F \text{ and } G$ are supported on the manifold $\mathcal{C}_d$ (see Theorem \ref{Modified Sufficiency}). Then testing $\mathcal{H}_{0}^{\prime}$ (see Equation \eqref{primary test}) is equivalent to testing the following hypothesis:
\begin{equation} \label{Secondary Hypothesis}
    \mathcal{H}_{0}: F = G \text{ vs. } \mathcal{H}_{1} : F \neq G.
\end{equation}
\end{t1} 

\section{Procedure: Test of Hypothesis} \label{Test Statistic}
This section presents a test statistic to perform a two-sample hypothesis test for $\mathcal{H}_0$ defined in Equation \eqref{Secondary Hypothesis}. The proposed test statistic is based on the energy distance (\cite{10.1214/23-EJS2203}) between the two distributions $F$ and $G$ supported on a $D$-dimensional compact smooth submanifold $\mathcal{M}$ of $\mathbb{R}^d$, $d \geq D$. Let $\rho$ be a metric on $\mathcal{M}$ and consider the random variables $X$, $X^\prime$, $Y$, $Y^\prime$ such that $X \overset{\mathcal{D}} {=} X^\prime$, and $Y \overset{\mathcal{D}} {=} Y^\prime$, where $X \sim F$, $Y \sim G$, and $U\overset{\mathcal{D}} {=} U^\prime$ indicate that the random variables $U$ and $U^\prime$ are identically distributed. Then, the energy distance between $F$ and $G$ denoted by $\mathcal{E}(F, G)$ is defined as:
\begin{equation} \label{energy distance}
 \mathcal{E}(F, G):= 2\mathbb{E}(\rho(X, Y)) - \mathbb{E}(\rho(X, X^\prime)) - \mathbb{E}(\rho(Y, Y^\prime)),    
\end{equation}
where $\mathbb{E} (U)$ denotes the expectation of a random variable $U$. 

The two-sample tests on Euclidean spaces based on the energy distance have been considered in the literature (see, e.g., \cite{InterStat}) and are shown to be consistent provided $\mathcal{E}(F, G)$ is a metric on the class of distributions under consideration. Recently, \cite{10.1214/23-EJS2203} extended the testing framework for the manifold-valued data and provided sufficient conditions for $\mathcal{E}(F, G)$ to be a metric. In the present context, the manifold under consideration is $\mathcal{C}_d$ (see Theorem \ref{Modified Sufficiency}) with the standard Euclidean metric. However, we require the following assumptions, for $\mathcal{C}_d$ to be a compact smooth submanifold of $\mathbb{R}^d$. 
\begin{A1} \label{A:1} $\mathcal{C}_d$ is a compact submanifold of $\mathbb{R}^d, d \geq 2$. 
\end{A1}
\begin{A1} \label{A:2} The class of probability distributions on the manifold $\mathcal{C}_d$ is defined as:
\begin{equation} \label{The class of distributions}
    \mathscr{C}:= \left\{F: F \text{ is absolutely continuous} \right\} .
\end{equation}
\end{A1}
Then, under Assumption \ref{A:1} and Assumption \ref{A:2}, the following proposition asserts that $\mathcal{E}(F, G)$ is a metric.
\begin{p1} \label{energy metric} $\mathcal{E}(F, G) = 0 $ if and only if $F = G$, for $F, G \in \mathscr{C}$. 
\end{p1}

We refer to Section \ref{Proofs} for the proof of Proposition \ref{energy metric}, which relies on the condition of strong negative type (\cite{StrongNegativetype}) for the metric space $\left(\mathcal{C}_d, \|.\|\right)$, where $\|.\|$ denotes the standard Euclidean metric in $\mathbb{R}^d, d \geq 2$. Now, we define the test statistic based on the energy statistic, which is the sample counterpart of $\mathcal{E}(F, G)$. 

Let $\mathcal{Z}:= \mathcal{S}_1 \cup \mathcal{S}_2$ denote the pooled sample obtained from the two samples defined in Theorem \ref{Modified Sufficiency}, that is, $\mathcal{S}_1: =\{\mathcal{V}(\mathscr{B}_{1}), \ldots, \mathcal{V}(\mathscr{B}_{n_1}) \}$ and $\mathcal{S}_2: =\{\mathcal{V}(\Tilde{\mathscr{B}}_1), \ldots, \mathcal{V}(\Tilde{\mathscr{B}}_{n_2}) \}$. Then, the energy statistic denoted by $\mathcal{E}_{n_1, n_2} (\mathcal{Z})$, is defined as  
\begin{equation} \label{Energy Statistic}
\mathcal{E}_{n_1, n_2} (\mathcal{Z}):=  \sum_{(X, Y) \in \mathcal{Z} \times \mathcal{Z}} 2(n_1 n_2)^{-1}\| X - Y\| -  \sum_{(X, X^\prime) \in \mathcal{S}_1 \times \mathcal{S}_1} n_1^{-2} \|X - X^\prime\| - \sum_{(Y, Y^\prime) \in \mathcal{S}_2 \times \mathcal{S}_2} n_2^{-2} \|Y - Y^\prime\|,
\end{equation}
where $A \times B$ denotes the Cartesian product of the two sets $A$ and $B$, and $\|.\|$ denotes the standard Euclidean metric in $\mathbb{R}^d, d\geq 2$. 

Then, we propose the following test for $\mathcal{H}_0$ (see Equation \eqref{Secondary Hypothesis}) based on $\mathcal{E}_{n_1, n_2} (\mathcal{Z})$. Let $\alpha \in (0, 1)$ be a fixed level of significance.  We propose to reject $\mathcal{H}_0$ at the level of significance $\alpha$, if the observed value $\mathcal{E}_{n_1, n_2} (\mathcal{Z}_{obs}) \geq C_{n_1, n_2}(\alpha)$, where $\mathcal{Z}_{obs}$ denotes the pooled sample containing observed values from the samples $\mathcal{S}_1$ and $\mathcal{S}_2$, and $C_{n_1, n_2}(\alpha)$ denotes the $(1-\alpha)$th quantile of the distribution of $\mathcal{E}_{n_1, n_2} (\mathcal{Z})$ under $\mathcal{H}_0$. 

To accomplish the proposed testing procedure, we compute $C_{n_1, n_2}(\alpha)$ using the permutation distribution of the test statistic $\mathcal{E}_{n_1, n_2} (\mathcal{Z})$ under $\mathcal{H}_0$. Note that, under $\mathcal{H}_0$, the random variables in $\mathcal{Z}$, say, $\mathcal{Z}:= \{Z_1, \ldots, Z_N\}$, where $N = (n_1 + n_2)$, are exchangeable. This implies that, under $\mathcal{H}_0$, any value of the test statistic across all $N!$ permutations of $\{Z_1, \ldots, Z_N\}$ is equally likely. Thus, under $\mathcal{H}_0$, $\mathcal{E}_{n_1, n_2} (\mathcal{Z}) \sim Unif\left\{\mathcal{E}_{n_1, n_2} (\mathcal{Z}_{obs}^\pi): \pi \in S_N \right\}$, where $\mathcal{Z}_{obs}^\pi$ denotes the observed pooled sample $\mathcal{Z}_{obs}$ with elements ordered according to the permutation $\pi$, and $S_N$ denotes the symmetric group on $\{1, \ldots, N\}$. This yields, under $\mathcal{H}_0$, for $k = \lceil (1- \alpha) N \rceil$, we have
\begin{equation} \label{Cutoff point}
    C_{n_1, n_2}(\alpha) = k\text{-}\min \{\mathcal{E}_{n_1, n_2} (\mathcal{Z}_{obs}^\pi): \pi \in S_N\},
\end{equation}
where $k$-$\min$(A) denotes the k-th smallest value from the set A.  

\subsection{Asymptotic Properties of Test} \label{Asymptotic Properties}
This subsection establishes the consistency of the proposed test. In other words, we show that the power of the proposed test tends to 1 as $\min(n_1, n_2) \xrightarrow{} \infty$. However, before performing an asymptotic analysis, we would like to highlight that the topological signatures of the observed random geometric objects are regularized for a suitable choice of $m \in \mathbb{N}$. That is, the observed random samples of barcodes lie in a regularized subset of the barcode space $\mathscr{B}_{\leq n}$, for a suitable choice of $m$. Therefore, recall that the proposed sufficient statistic $\mathcal{V}$ is a measurable transformation from the regularized barcode space $\mathscr{B}_{\leq n}^m$ to $\mathcal{C}_{d}$ (see Theorem \ref{Modified Sufficiency}). A data-driven choice of $m$ would vary with sample size, rendering the domain of $\mathcal{V}$ sample-dependent, and thereby complicating a rigorous asymptotic analysis. 

To remedy this, we propose a universal value of $m$ that can be used to regularize the barcode space. First, recall from the definition of $\mathscr{B}_{\leq n}^m$ (see Equation \eqref{regularized barcode space}), we subset only those barcodes that consist of topological features that satisfy the following for $m \in \mathbb{N}$:
\begin{equation} \label{choice of m}
    b_{i} \leq m (d_{i} -b_{i}) \implies b_{i} \leq \frac{m}{ m +1} d_{i} \text{ for all } i = 1, \ldots, n,
\end{equation}
where $d_i$ is the death time of the $i$-th feature in the barcode, and is related to the persistence $\ell_i$ as, $\ell_i = d_i - b_i$. 
The condition in Equation \eqref{choice of m} can be interpreted as choosing features from a subset of persistent diagrams depending on $m$. Note that in a typical persistence diagram, the birth always precedes the death of a topological feature. Therefore, all the topological features of a persistent diagram are in the region $\{(b, d) \in \mathbb{R}^2: 0 < b \leq d\}$. The condition in Equation \eqref{choice of m} reduces this region by scaling the death times by $m / m+1$. Thus, the higher the value of $m$, the wider the regularized region, which encompasses the features close to the diagonal in regularized subsets of barcodes. Therefore, if we choose a smaller value of $m$, say $m = 1$, then we will leave out most of the features that are close to the diagonal, while if we choose a higher value of $m$, say $m = 100$, then the region $\{(b,d): b \leq 0.99 d\}$ is closer to the region $\{(b, d): b \leq d \}$. Thus, a suitable larger value, say $m = 100$ allows us to subset most of the features from the persistence diagram. Hence, an appropriate and universal choice of $m$ to draw random samples from $\mathscr{B}_{\leq n}^{m}$ could be $m = 100$. 


Now, we state the consistency of the proposed permutation test in the following theorem.

\begin{t1} \label{Consistency of the test} Let the sample sizes $n_1$ and $n_2$ are such that $n_1/(n_1+n_2) \xrightarrow{} \lambda \in (0,1) $ as $\min(n_1, n_2) \xrightarrow{} \infty$. Then under Assumption \ref{A:1} and Assumption \ref{A:2}, the test based on $\mathcal{E}_{n_1, n_2}(\mathcal{Z})$ for $\mathcal{H}_{0}$ (see Equation \eqref{Secondary Hypothesis}), is consistent, that is, for the following probability under $\mathcal{H}_1$, we have 
$$ \mathbb{P}_{\mathcal{H}_1} \left(\mathcal{E}_{n_1, n_2}(\mathcal{Z}) \geq C_{n_1,n_2}(\alpha)\right) \xrightarrow{} 1 \text{ as } \min (n_1, n_2) \xrightarrow{} \infty.$$
\end{t1}

\section{Conclusion} \label{Conclusion}
We propose a two-sample testing framework to detect topological differences in random geometric objects. In the course of this study, we propose a sufficient statistic derived from tropical embeddings of barcodes to place the testing framework on a standard statistical footing. As an application of Theorem \ref{Modified Sufficiency}, we establish that it is equivalent to performing a two-sample test on the barcode space to a two-sample problem on the ordered convex cone ($\mathcal{C}_d$) in $\mathbb{R}^d$ (see Theorem \ref{Equivalence of Hypothesis}). We propose a two-sample test on the manifold $\mathcal{C}_d$ based on the manifold energy statistics and derive its consistency. The proposed testing framework is a generalized framework of hypothesis testing framework proposed by \cite{Robinson2017} and \cite{Blumberg2014}. In particular, the proposed testing framework can be adapted for the ensembles of point cloud data. Moreover, the proposed testing framework provides an alternative to the testing framework proposed by \cite{RandomShapes2025}. As a future consideration, it would be tempting to explore the possibility of extending the proposed framework for a time series of random geometric objects.

\section{Appendix} \label{Appendix}
\subsection{Definitions}
\begin{d1} \label{Tame set} (\textbf{Tame Set}) We use the notion of o-minimal structures from \cite{CurryMukherjeeTurner2022} to define tame sets. Let $\mathcal{P}(\mathbb{R}^d)$ denote the power set of $\mathbb{R}^d$, and $A \times B$ denotes the Cartesian product of two sets $A$ and $B$. An o-minimal structure is defined as $\mathcal{O} := \left\{\mathcal{O}_d: d \geq 1\right\}$, where $\mathcal{O}_d \subseteq \mathcal{P}(\mathbb{R}^d)$ satisfying the following conditions:  
\begin{enumerate}
    \item Sets in $\mathcal{O}_d$ are closed under finite intersection and complement.

    \item For any set $A \in \mathcal{O}_d$, we have $A \times \mathbb{R} \in  \mathcal{O}_{d + 1} $ and $\mathbb{R} \times A \in  \mathcal{O}_{d + 1} $.

    \item Let $\pi : \mathbb{R}^{d + 1} \xrightarrow{} \mathbb{R}^d $ be an  axis-aligned projection map. Then for any for any set $A \in \mathcal{O}_{d +1}$, we have $\pi(A) \in \mathcal{O}_d $.  

    \item $\mathcal{O}$ is closed with respect to all the operations of $\mathbb{R}$ that make it an ordered field, that is, the operations like comparison ($<$), addition, and multiplication. 

    \item The only sets in $\mathcal{O}_1$ are all finite unions of points and open intervals of $\mathbb{R}$. 
\end{enumerate}
Then, the elements of $\mathcal{O}$ are called \textbf{tame sets}. 
\end{d1}

\begin{d1} \label{Gromov–Hausdorff distance.} (\textbf{Gromov–Hausdorff distance}) 
We define the Gromov–Hausdorff distance between two metric spaces $(X, d_X)$ and $(Y, d_Y)$ in terms of correspondences as in  \cite{RandomGeometricObjectAOS2465}. A correspondence is a subset $C \subset X \times Y $ that satisfies the following:
\begin{enumerate}
    \item For all $x \in X$, $\exists y \in Y$ such that $(x, y) \in C$ 

    \item For all $y \in y$, $\exists x \in X$ such that $(x, y) \in C$. 
\end{enumerate}
The distortion of the correspondence $C$ is defined as:
$$ dist(C) = \sup_{(x_1,x_2), (y_1, y_2) \in C}  \left| d_X(x_1, x_2) - d_Y(y_1, y_2)\right|.$$

Then, the Gromov–Hausdorff distance $d_{GH}(X, Y)$ between the metric spaces $X$ and $Y$ is defined as:
$$ d_{GH}(X, Y) = \frac{1}{2} \inf \left\{ dist(C) : C \in \mathscr{C}\right\},$$
where $\mathscr{C}$ denotes the class of all correspondences between $X$ and $Y$. 
\end{d1}

\begin{d1} \label{Bottleneck distance} \textbf{(Bottleneck distance)} \cite{Carlsson(2014)} Let $\mathscr{B}_{1}$ and $\mathscr{B}_2$ be two barcodes in $\mathscr{B}_{\leq n}$ (see Equation \eqref{Barcode space}). This implies that $\mathscr{B}_1$ and $\mathscr{B}_2$ can be written as finite collections of intervals. That is, $\mathscr{B}_1:= \{I_i: i \in [N]\}$ and $\mathscr{B}_2:= \{J_i: i \in [M]\}$, for some positive integer $N$ and $M$ such that $\max(N, M) \leq n$. Recall that, here $[n]$ represent the set $\{1, \ldots, n\}$ for any $n \in \mathbb{N}$. Now, to define the bottleneck distance, we first need to specify the distance between two features in a barcode as well as the distance between a feature and the diagonal $\Delta = \{[b, b): b \geq 0\}$ containing bars of length 0. We define the distance between two features $I:= [b_1, \ell_1)$ and $J:= [b_2, \ell_2)$ as:
$$\updelta_{\infty}(I, J):= \max \left( |b_1 - b_2|, |(b_1 + \ell_1) - (b_2 + \ell_2)|\right),$$
where $b_i$ represents the birth time and $\ell_i$ represents the persistence of the $i$-th feature, $i= 1, 2$.  
The distance between a feature $[b, \ell)$ and the diagonal $\Delta$ is defined as:
$$ \updelta_{\infty}([b, \ell), \Delta):= \frac{\ell}{2}.$$

Now, consider a bijection $\upphi : A \xrightarrow{} B$, where $A \subseteq [N] $ and $B \subseteq [M]$, and define the penalty $\uprho(\upphi)$ of $\upphi$ as:
$$ \uprho(\upphi):= \max \left( \max_{i \in A} \left(\updelta_{\infty}\left(I_i, J_{\upphi(i)}\right)\right), \max_{i \in [N] \setminus A} \left( \updelta_{\infty}\left(I_i, \Delta \right)\right), \max_{i \in [M] \setminus B} \left( \updelta_{\infty}\left(J_i, \Delta \right)\right)\right)$$
Then the bottleneck distance between $\mathscr{B}_1$ and $\mathscr{B}_2$ is denoted by $\delta_B(\mathscr{B}_1, \mathscr{B}_2)$, and defined as:
$$ \delta_B(\mathscr{B}_1, \mathscr{B}_2):= \min_{\upphi} \left(\uprho(\upphi)\right).$$
\end{d1}

\begin{d1} \label{General Factorization Theorem} (\textbf{General Factorization Theorem \cite{Bahadur1954}})
\noindent Let $(\mathfrak{X}, \mathfrak{F})$ be a measurable space with a family of probability measures $\mathfrak{M}$ dominated by a $\sigma$-finite measure $\lambda$. Then a statistic $T$ is sufficient for $\mathfrak{M}$ if and only if there exist a non-negative measurable function $h$ on $\mathfrak{X}$ and a set of non-negative measurable functions $\{g_{\vartheta}: \vartheta \in \mathfrak{M}\}$ on the range of $T$ such that for each $\vartheta \in \mathfrak{M}$, the Radon-Nikodym derivative $f_{\vartheta} \equiv d\vartheta / d\lambda$ admits the factorization $$f_{\vartheta}(x) = h(x)g_{\vartheta}(T(x)), x \in \mathfrak{X}.$$
\end{d1}

\subsection{Proofs of Theorems and Propositions} \label{Proofs}

\begin{proof} [\textbf{Proof of Theorem \ref{Modified Sufficiency}.}] 
We proceed by establishing that for a barcode $\mathscr{B} \sim\vartheta \in \mathcal{P}$, the Radon-Nikodym derivative $f_\vartheta \equiv d\vartheta/d\lambda$ factors as:
$$f_\vartheta(\mathscr{B}) = h(\mathscr{B}) g_\vartheta \left(\mathcal{V} (\mathscr{B})\right),$$
where $h$ is a non-negative measurable function on $\mathscr{B}_{\leq n}^m$ and $g_\vartheta$ is a non-negative measurable function on $\mathcal{C}_d$. Recall that $\mathcal{C}_d:=\{(x_1, \ldots, x_d)^\top \in \mathbb{R}^d: x_1\leq x_2\leq \ldots, \leq x_d \}$. Then by the general factorization theorem (see Definition \ref{General Factorization Theorem}), the map $\mathcal{V}$ will be a sufficient statistic.

First, we use the property that the map $\mathcal{V}: \mathscr{B}_{\leq n}^m \xrightarrow{} \mathcal{C}_d$ is injective. This is because the tropical coordinates defined in Equation \eqref{Finite Coordinates} separate the barcodes in $\mathscr{B}_{\leq n}^m$ by applying Preposition 2.8 of \cite{Monod2019} and Theorem \ref{Tropical coordinates}  (see \cite{Monod2019}). This implies that for any two distinct point $\mathscr{B}_1$ and $\mathscr{B}_2$ in $\mathscr{B}_{\leq n}^m$, we have $\mathcal{V}(\mathscr{B}_1) \neq \mathcal{V}(\mathscr{B}_2)$. Consequently, $\mathcal{V}$ is an embedding, therefore, there exists a function $\eta$ such that $\eta \circ \mathcal{V}$ and $ \mathcal{V} \circ \eta$ are identity maps in $\mathscr{B}_{\leq n}^m$ and $\mathcal{C}_d$, respectively. Thus, we can write $f_\vartheta(\mathscr{B}) = h(\mathscr{B})g_{\vartheta}(\mathcal{V}(\mathscr{B}))$, for $h(\mathscr{B}) = 1$ and $g_{\vartheta} \equiv f_\vartheta \circ \eta$. It is evident that both $h$ and $g_{\vartheta}$ are non-negative, $g_{\vartheta}$ is non-negative, as $f_\vartheta$ is a probability density on $\mathscr{B}_{\leq n}^m$. Now, we verify the measurability of the maps $h$ and $g_\vartheta$ to apply the general factorization theorem (see Definition \ref{General Factorization Theorem}).  

Note that the map $h(\mathscr{B}) = 1$ is a constant map, hence it is continuous. Therefore, by Theorem 1.5 of \cite{Parthasarathy}, $h$ is measurable. Next, to show that $g_{\vartheta} \equiv f_{\vartheta} \circ \eta$ is measurable, we use the fact that the composition of two measurable maps is measurable. Therefore, we need to show that $\eta$ is measurable as $f_{\vartheta}$ is measurable by the Radon-Nikodym theorem. We use the Kuratowski theorem (see Chapter 3 in \cite{Parthasarathy}), which states that the inverse of an injective, measurable map between complete and separable metric spaces is measurable. Now, since both the metric spaces  $(\mathscr{B}_{\leq n}^m, \delta_B )$ and $(\mathcal{C}_d, \delta)$ are closed subspaces of complete and separable metric spaces $(\mathscr{B}_{\leq n}, \delta_B)$ and $(\mathbb{R}^d, \delta)$, respectively, $\delta$ denotes the standard Euclidean metric in $\mathbb{R}^d$. This implies that both the metric spaces $(\mathscr{B}_{\leq n}^m, \delta_B )$ and $(\mathcal{C}_d, \delta)$ are complete and separable. We refer to Theorem 3.2 of \cite{Blumberg2014} for completeness and separability of $(\mathscr{B}_{\leq n}, \delta_B)$. Consequently, the inverse of $\mathcal{V}$, that is, the map $\eta$ is measurable. Hence, the embedding $\mathcal{V}$ is a sufficient statistic for $\mathcal{P}$ by the general factorization theorem (see Definition \ref{General Factorization Theorem}). This completes the proof of Theorem \ref{Modified Sufficiency}.
\end{proof}

\begin{proof} [\textbf{Proof of Theorem \ref{Equivalence of Hypothesis}}] We use Theorem \ref{Modified Sufficiency} and apply the general factorization theorem (see Definition \ref{General Factorization Theorem}) for $h(\mathscr{B})\equiv 1$ without loss of generality to establish Theorem \ref{Equivalence of Hypothesis}. Let $f_{\mu}$, $f_{\vartheta}$, $g_{\mu}$ and $g_{\vartheta}$ denote the probability densities of $\mu$, $\vartheta$, $F$ and $G$, respectively. Recall that, $F$ and $G$ are the probability distributions corresponding to the induced probability measures $\mu \circ \mathcal{V}^{-1}$ and $\vartheta \circ \mathcal{V}^{-1}$, respectively. 

We need to show that any statistical decision (accept or reject) for $\mathcal{H}_{0}^\prime$ (see, Equation \eqref{primary test}) is valid for $\mathcal{H}_{0}$ (see, Equation \eqref{Secondary Hypothesis}), and vice versa. Consider the situation when we accept the null hypothesis $\mathcal{H}_{0}^\prime$. This implies that for every $A \in \sigma(\mathscr{B}_{\leq n}^m)$ and $\mathscr{B} \in \mathscr{B}_{\leq n}^m$, we have:
\begin{equation} \label{Eq4.2}
    \mu(A) = \nu(A) \Longleftrightarrow f_{\mu}(\mathscr{B}) = f_{\nu}(\mathscr{B}),
\end{equation}
where Equation \eqref{Eq4.2} follows from the fact that the probability density of a random variable uniquely characterizes its probability measure. Now, using the sufficiency of the tropical embedding $\mathcal{V}$ from Theorem \ref{Modified Sufficiency}, we have:
\begin{equation} \label{Eq4.3}
   f_{\mu}(\mathscr{B}) = f_{\nu}(\mathscr{B}) \Longleftrightarrow g_{\mu}(\mathcal{V}(\mathscr{B})) = g_{\nu}(\mathcal{V}(\mathscr{B})) \Longleftrightarrow F(\mathcal{V}(\mathscr{B})) = G(\mathcal{V}(\mathscr{B})). 
\end{equation}
Therefore, by Equation \eqref{Eq4.2} and \eqref{Eq4.3}, for any arbitrary $\mathscr{B} \in \mathscr{B}_{\leq n}^m$ and $A \in \sigma(\mathscr{B}_{\leq n}^m)$, we have:
\begin{equation} \label{Accept}
    \mu(A) = \nu(A) \Longleftrightarrow  F(\mathcal{V}(\mathscr{B})) = G(\mathcal{V}(\mathscr{B})).
\end{equation}

Now, consider the situation when we reject $\mathcal{H}_{0}^\prime$, that is, there exists $A \in \sigma(\mathscr{B}_{\leq n}^m)$ and  $\mathscr{B} \in \mathscr{B}_{\leq n}^m$ such that:
\begin{equation} \label{Eq:4.4}
    \mu(A) \neq \nu(A) \Longleftrightarrow f_{\mu}(\mathscr{B}) \neq f_{\nu}(\mathscr{B}).
\end{equation}
This further implies by Theorem \ref{Modified Sufficiency} that:
\begin{equation} \label{Eq4.5}
    f_{\mu}(\mathscr{B}) \neq f_{\nu}(\mathscr{B}) \Longleftrightarrow g_{\mu}(\mathcal{V}(\mathscr{B})) \neq  g_{\nu}(\mathcal{V}(\mathscr{B})) \Longleftrightarrow F(\mathcal{V}(\mathscr{B})) \neq  G(\mathcal{V}(\mathscr{B})).
\end{equation}
Therefore, by Equation \eqref{Eq:4.4} and \eqref{Eq4.5}, there exists $A \in \sigma(\mathscr{B}_{\leq n}^m)$ and $\mathscr{B} \in \mathscr{B}_{\leq n}^m$ such that:
\begin{equation} \label{Reject}
    \mu(A) \neq \nu(A) \Longleftrightarrow  F(\mathcal{V}(\mathscr{B})) \neq G(\mathcal{V}(\mathscr{B})) .
\end{equation}
Thus, using Equation \eqref{Accept} and \eqref{Reject} it is established that both the hypothesis $\mathcal{H}_{0}^\prime$ (see, Equation \eqref{primary test}) and $\mathcal{H}_{0}$ (see, Equation \eqref{Secondary Hypothesis}) are equivalent. This establishes the statement in Theorem \ref{Equivalence of Hypothesis}. 
\end{proof}

\begin{proof} [\textbf{Proof of Proposition \ref{energy metric}}] We need to show that the metric space $\left(\mathcal{C}_d, \|.\|\right)$ has strong negative type (\cite{StrongNegativetype}). Then, by Proposition 3 of \cite{SzekelyRizzo2017Energy}, $\mathcal{E}(F, G)$ will be a metric on the class of distribution functions $\mathscr{C}$ (see Equation \eqref{The class of distributions}). The metric space $\left(\mathcal{C}_d, \|.\|\right)$ has strong negative type if for any two probability distributions $F$ and $G$ supported on $\mathcal{C}_d$ and for the random variables $X$, $X^\prime$, $Y$ and $Y^\prime$ such that $X \overset{\mathcal{D}}{=} X^\prime$ and $Y \overset{\mathcal{D}}{=}Y^\prime$, where $X \sim F$ and $Y \sim G$, we have:
\begin{equation} \label{Strong Negative Type Condition}
    2 \mathbb{E}\|X - Y\| - \mathbb{E}\|X - X^\prime\| - \mathbb{E}\|Y - Y^\prime\| \geq 0,
\end{equation}
 such that equality is attained in Equation \eqref{Strong Negative Type Condition} if and only if $F = G$. The condition of strong negative type for the manifold $\mathcal{C}_d$ holds under the Euclidean metric by Theorem 2.1 of \cite{JMA}. This implies that the metric space $\left(\mathcal{C}_d, \|.\|\right)$ has strong negative type. Hence, by Proposition 3 of \cite{SzekelyRizzo2017Energy}, $\mathcal{E}(F, G)$ is metric on $\mathscr{C}$. This establishes the assertion in Proposition \ref{energy metric}.           
\end{proof}

\begin{proof} [\textbf{Proof of Theorem \ref{energy metric}}] The proof proceeds along the following two steps. First, we show that the energy statistic $\mathcal{E}_{n_1, n_2}(\mathcal{Z})$ converges in probability to its population counterpart $\mathcal{E}(F,G)$ under $\mathcal{H}_1$, that is, we have: 
\begin{equation} \label{U-statistics}
    \mathcal{E}_{n_1, n_2}(\mathcal{Z}) \xrightarrow{P} \mathcal{E}(F,G) \text{ as } \min(n_1, n_2) \xrightarrow{} \infty
\end{equation}
The equation \eqref{U-statistics} follows directly from the application of the asymptotics of U-statistics. In particular, we apply Theorem 12.6 of \cite{vanderVaart1998} to the first term of $\mathcal{E}_{n_1, n_2}(\mathcal{Z})$ and Theorem 12.3 of \cite{vanderVaart1998} to the remaining two terms of $\mathcal{E}_{n_1, n_2}(\mathcal{Z})$.  

Second, we use Lemma A.2 of \cite{10.1214/23-EJS2203}, which establishes that for every $\epsilon > 0$ there exists $ 0 < M < \infty$ such that for any permutation $\pi \in S_N$, we have: 
\begin{equation} \label{Critical Point Asymptotics}
\liminf_{\min (n_1, n_2) \xrightarrow{} \infty}\mathbb{P} \left( (n_1 + n_2)\mathcal{E}_{n_1, n_2}( \mathcal{Z}^\pi) < M\right) \geq 1 - \epsilon,
\end{equation}
 where $\mathcal{Z}^\pi$ denotes the pooled sample with elements ordered according to the permutation $\pi$ on $\{1, \ldots, (n_1 + n_2)\}$. 

 Now, consider the following probability under $\mathcal{H}_1$:
 \begin{align*}
     \mathbb{P}_{\mathcal{H}_1} \left(\mathcal{E}_{n_1,n_2}(\mathcal{Z}) \geq C_{n_1, n_2}(\alpha)\right) &= \mathbb{P}_{\mathcal{H}_1} \left( (n_1 + n_2)\mathcal{E}_{n_1,n_2}(\mathcal{Z}) \geq (n_1 + n_2)C_{n_1, n_2}(\alpha)\right)\\ 
     &\geq \mathbb{P}_{\mathcal{H}_1} \left( (n_1 + n_2)\mathcal{E}_{n_1,n_2}(\mathcal{Z}) \geq M\right) \tag{E.1} \label{E.1}\\
     &\xrightarrow{} \mathbb{P}_{\mathcal{H}_1} \left( \mathcal{E}(F, G) \geq 0\right) \text{ as } \min(n_1, n_2) \xrightarrow{} \infty \tag{E.2} \label{E.2}\\
     &= 1, \text{ as } \mathcal{E}(F, G) > 0, \text{ by Preposition \ref{energy metric} under } \mathcal{H}_1, 
 \end{align*}
 where \eqref{E.1} follows from the application of Equation \eqref{Critical Point Asymptotics} to Equation \eqref{Cutoff point} and \eqref{E.2} follows from Equation \eqref{U-statistics}. This establishes the consistency of the proposed test.   
\end{proof}
\bibliographystyle{apalike} 
\bibliography{TDA}
\end{document}